\documentclass[]{elsarticle}

\usepackage{color}
\usepackage[margin=1in, letterpaper]{geometry}
 
\usepackage{amsmath, amssymb} 
\newcommand{\bm}[1]{{\mbox{\boldmath $#1$}}} 
\usepackage{float} 
\usepackage{ulem} 
\newcommand{\argmin}{\mathop{\rm arg~min}\limits} 

\biboptions{authoryear}

\begin{document}

\begin{frontmatter}

\title{Chaos may enhance expressivity in cerebellar granular layer}

\author[mymainaddress]{Keita Tokuda\corref{mycorrespondingauthor}}
\cortext[mycorrespondingauthor]{Keita Tokuda}
\ead{tokudakeita-tky@umin.ac.jp}
\address[mymainaddress]{
Department of Computer Science, University of Tsukuba, 1-1-1 Tennodai, Tsukuba, Ibaraki 305-8577, Japan
}

\author[NFaddress,seiken,csis]{Naoya Fujiwara}
\address[NFaddress]{Graduate School of Information Sciences, Tohoku University, 6-3-09 Aoba, Aramaki-aza Aoba-ku, Sendai, Miyagi, 980-8579, Japan}

\address[seiken]{Institute of Industrial Science, The University of Tokyo, 4-6-1 Komaba, Meguro-ku, Tokyo 153-8505, Japan}

\address[csis]{Center for Spatial Information Science, The University of Tokyo, 5-1-5 Kashiwanoha, Kashiwa-shi, Chiba 277-8568, Japan}

\author[ASaddress]{Akihito Sudo}
\address[ASaddress]{Faculty of Informatics, Shizuoka University, 3-5-1 Johoku, Naka-ku, Hamamatsu-shi, Shizuoka, Japan}

\author[YKaddress,seiken]{Yuichi Katori}
\address[YKaddress]{The School of Systems Information Science, Future University Hakodate, 116-2 Kamadanakano-cho, Hakodate, Hokkaido 041-8655, Japan}

\begin{abstract}
Recent evidence suggests that Golgi cells in the cerebellar granular layer are densely connected to each other with massive gap junctions. Here, we propose that the massive gap junctions between the Golgi cells contribute to the representational complexity of the granular layer of the cerebellum by inducing chaotic dynamics. We construct a model of cerebellar granular layer with diffusion coupling through gap junctions between the Golgi cells, and evaluate the representational capability of the network with the reservoir computing framework. First, we show that the chaotic dynamics induced by diffusion coupling results in complex output patterns containing a wide range of frequency components. Second, the long non-recursive time series of the reservoir represents the passage of time from an external input. These properties of the reservoir enable mapping different spatial inputs into different temporal patterns.
\end{abstract}


\end{frontmatter}


\section{Introduction}
\noindent
Recent experimental studies have revealed that neighboring Golgi cells in the granular layer of the cerebellar cortex are densely interconnected with gap junctions that allow direct diffusion of ions between neuronal intracellular spaces (\cite{Dugue2009,Vervaeke2010}). \cite{Vervaeke2010} reported that more than 80~\% of neighboring neuron pairs are interconnected with gap junctions, and that each Golgi cell is connected to approximately 10 other Golgi cells via gap junctions. They also showed that the diffusion current between neighboring Golgi cells has the effect of transiently desynchronizing the spike activities after external excitation (\cite{Vervaeke2010}). This is contradictory to the classical view of the role of gap junctions, which is to synchronize nearby neurons (\cite{Watanabe}). In spite of the complex effect of diffusion coupling between Golgi cells on the ongoing dynamics, the causal relationship between this dynamics and cerebellar computation has yet to be elucidated.

Several theoretical studies have pointed out that diffusion coupling between nonlinear oscillators not necessarily realizes synchronization, but also induces instability (\cite{Turing}) or even chaotic activity (\cite{Kuramoto, FUJII2004151,Tsuda2004, Schweighofer4655, TOKUDA2010836, Katori2010,Tadokoro2011, Tokuda2019}). \cite{FUJII2004151} and \cite{Tsuda2004} reported that introducing diffusion coupling through gap junctions between class 1 neurons induces chaotic dynamics. 
\cite{Schweighofer4655} proposed a theory that the abundant gap junctions in the inferior olive produce chaotic neural activity that enables efficient transmission of information in the high-frequency components of inputs. 
It has also been proposed that the adaptive strength of the gap junction in the inferior olive regulates the degrees of freedom of the system, and the brain modifies the gap junction strength during learning to ensure that the system operates at an optimal level of degrees of freedom (\cite{KAWATO2011791, TOKUDA201342, TOKUDA201758, Hoang542183}). The possible computational role of diffusion coupling through gap junction in the granular layer should be elucidated as well.

The majority of cerebellar computational theories assume that the cerebellum is a supervised machine that learns a desirable input-output relationship (\cite{Marr, ALBUS197125,  Ito1970, Kawato1987, Buonomano1994, Wolpert1998, Schweighofer4655, YAMAZAKI2007290, Raymond}). It is well known that two major input pathways converge on the Purkinje cells; the mossy fiber-granular layer-Purkinje cell pathway originating from a precerebellar nucleus such as the pontine nucleus, and the climbing fiber-Purkinje cell pathway originating from the inferior olive (\cite{Paxinos_cerebellum}). These theories assume that the former pathway is the input layer of the supervised machine and the latter pathway conveys the supervising signals. The computational role of the cerebellar granular layer is assumed to be the preprocessing -- feature engineering -- of incoming signals from the mossy fibers. It transforms an input to a dynamical representation in a high-dimensional space realized by the enormous number ($\sim 10^{11}$ in human) of the granular neurons (\cite{Marr, ALBUS197125, Badura2017,Raymond}). The granule cells and Golgi cells are the two major components of the cerebellar granular layer. Even though the major outputs of the granular layer are conveyed by the parallel fibers of the granule cells, the Golgi cells are also thought to play the central role of forming the representation because of the lack of direct recurrent connection within granule cells (\cite{Marr, ALBUS197125, Raymond}).

It has long been known that the cerebellum plays a crucial role in motor learning, which requires execution of sequential movements with temporally precise timing (\cite{ItoM}). For example, a vast amount of experimental studies have characterized the essential information flow in the cerebellum that supports motor learning called classical eyeblink conditioning (\cite{Thompson}). In a typical eyeblink conditioning, the animal is exposed to paired presentation of tone stimulus and a periorbital air puff stimulus intervened with a fixed interval (typically 250 ms) repetitively. After learning occurs, the animal acquires a temporally precise motor response (eyeblink) to the tone. The eyeblink response is precisely timed at the air puff onset with millisecond precision. It is also known that the interval discrimination task can be learned in eyeblink conditioning: animals can learn to elicit eyeblink responses at different latencies to different tone stimuli (\cite{Kehoe1993, Green01052005}). Vast amount of evidence supports the fact that the cerebellum acquires the desired map to return a specific spatiotemporal pattern to a specific input. To explain this computation of the cerebellum, \cite{Buonomano1994} proposed a model of the granular layer consisting of sparse reciprocally connected granule cells and Golgi cells that are capable of representing the passage of time from the onset of an external sensory stimulus. In their model, mossy fiber excitation conveying the information of external tone stimulus elicits activity of the granule cells and the Golgi cells, with different sub-populations activated at different times. As a result, a specific sub-population of the granule cells is activated at a specific time from the onset of the stimulus, thereby representing the passage of time. This model successfully explains an important aspect of the behavioral and physiological traits in eyeblink conditioning.

\cite{YAMAZAKI2007290} extended Buonomano's model, and proposed the view that the cerebellum is a liquid state machine -- a type of reservoir machine (\cite{jaeger:techreport2001, Maass2002}). In the reservoir computing framework, the input signals to the system project to a recurrent network called reservoir that has a highly nonlinear dynamics in a high dimensional space, and only the readout connections from this reservoir are modified to give the desired output signals. In Yamazaki's model, the general computational role of the cerebellum is to acquire a map between spatiotemporal input patterns and desired spatiotemporal output patterns. This is a natural elaboration of the classical Marr-Albus-Ito model regarding the cerebellum as a supervised machine, in that the reservoir machine can process spatiotemporal patterns. The granular layer works as the reservoir, and long-term depression (LTD) of the parallel fiber-Purkinje cell connection works as the learning rule. The Purkinje cell corresponds to the output neuron of the reservoir. They showed that the network of random recurrent connections between granule and Golgi cells realized temporally specific activation of different sub-populations of granule cells in response to external inputs. Their model successfully acquires a function that maps specific different inputs to specific different temporal patterns. To date, several studies of cerebellar function with the reservoir computing framework have been conducted (\cite{Yamazaki2012, RossertEdgeOfChaos}). However, even though these theories assume the inevitable functional role of the Golgi cells in realizing the reservoir, to our knowledge, no study has focused on the functional role of massive gap junctions between the Golgi cells. Considering the facts that chaotic activity is related to the performance of the reservoir (\cite{jaeger:techreport2001, EdgeOfChaos2, EdgeOfChaos, Sussillo2009, YILDIZ20121, Laje2013}) and that gap junction often induces chaotic activity (\cite{Kuramoto, FUJII2004151,Tsuda2004, Schweighofer4655, Katori2010, TOKUDA2010836, Tadokoro2011, Tokuda2019}), it is necessary to elucidate how the gap junctions affect the computational performance of the granular layer as the reservoir, especially in terms of the effect of chaotic dynamics it may produce.
\section{Methods}
\begin{figure}[h]
\begin{center}
  \includegraphics[]{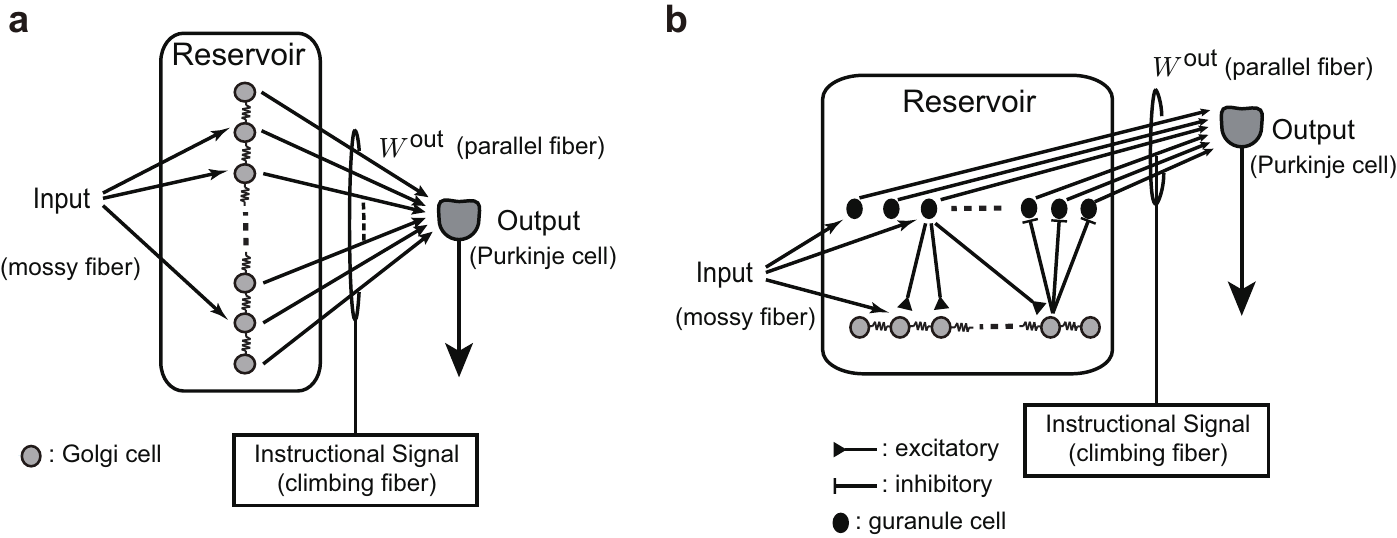}
    \caption{Two network architectures of the reservoir machine considered in the current study. The assumed corresponding anatomical structures of the cerebellum are also indicated. (a) The Golgi cells coupled with gap junctions are the reservoir. Other structures in the granular layer, such as the granule cells and chemical synapses, are omitted in order to focus on studying the computational performance of neurons with gap junctions acting as the reservoir. (b) The model incorporating the granule cells, the excitatory projections from the granule cells to the Golgi cells, and the inhibitory projections from the Golgi cells to the granule cells.}
 \label{arch}
    \end{center}
\end{figure}
\subsection{The network architecture}
Figure~\ref{arch} shows the schematic diagrams of the two network architectures of the reservoir machines studied in the current study. We take the view that the cerebellum is a reservoir machine (\cite{YAMAZAKI2007290}). In this view, the granular layer of the cerebellum works as the reservoir, the mossy fiber projecting to the granular layer is the input, and the Purkinje cell is the output neuron. Learning via synaptic modification is realized by changing the connection strength of the parallel fibers, which are the readout connections. The reservoir maps an incoming input into a high-dimensional time series by nonlinear dynamics. Figure~\ref{arch}(a) shows a simple reservoir machine composed of only Golgi cells that are mutually connected with diffusion coupling through gap junctions. In this study, we mainly focused on this model to evaluate the computational performance of diffusion-induced chaos as a reservoir machine. To confirm whether the results obtained with this model can also be observed in a more realistic situation, we perform a simulation incorporating the granule cells as well (Fig.~\ref{arch}(b)). This model incorporates other major components of the granular layer: the granule cells, the excitatory projections from the granule cells to the Golgi cells, and the inhibitory projections from the Golgi cells to the granule cells. The readout projection to the Purkinje cell originates from the granule cells as the actual cerebellar anatomical structure. Note that we do not incorporate a feedback loop from the reservoir output to the input in this study.
\subsection{The model dynamics}
A cerebellar Golgi cell is known to have the following properties: (1) it shows a periodic activity {\it in vitro} (\cite{Forti, Solinas, Vervaeke2010}), (2) its spike frequency increases as external input increases (\cite{Forti, Solinas}), and (3) its diffusion coupling induces desynchronization of neighboring cells (\cite{Vervaeke2010}). We model the Golgi cells with the $\mu$-model, which is a simple model described by a two-dimensional ordinary differential equation (\cite{Tsuda2004}). The $\mu$-model is a class 1 neuron model that shows spiking activity after a saddle-node bifurcation occurs as tonic external input increases. This model shows periodic activity under isolated conditions with a tonic input, increases spike frequency to increasing tonic input, and shows aperiodic activity when coupled with diffusion (\cite{Tsuda2004, Katori2010, Tadokoro2011, TOKUDA201342}; \cite{Tokuda2019}). The simple model shown in Fig.~\ref{arch}(a) is described as follows:
\begin{eqnarray}
\frac{\mathrm{d}V^{\mathrm{go}}_i}{\mathrm{d}t} 
&=&
- R_i^{\mathrm{go}} 
- \mu \cdot \left(V^{\mathrm{go}}_i \right) ^2(V^{\mathrm{go}}_i \mathrm{-\frac{3}{2}})
+ J^{\mathrm{go}}_i
+I^{\mathrm{go, tonic}} 
+I_{i}^{\mathrm{go}, \mathrm{input}}
,
 \label{V_i}
\\ 
\frac{\mathrm{d}R^{\mathrm{go}}_{i}}{\mathrm{d}t} &=& - R_i^{\mathrm{go}} - \mu \cdot \left( V^{\mathrm{go}}_{i} \right)^2,
 \label{R_i}
\\
J^{\mathrm{go}}_i&=& \left\{ \begin{array}{ll} 
g_{\mathrm{GJ}}(V^{\mathrm{go}}_{2}-V^{\mathrm{go}}_{1})
 & \ (\mathrm{for} \quad i = 1), \\
g_{\mathrm{GJ}}(V^{\mathrm{go}}_{i+1}+V^{\mathrm{go}}_{i-1}-2V^{\mathrm{go}}_{i})
 & \ (\mathrm{for} \quad i = 2,...,N_{\mathrm{go}}-1),\\
g_{\mathrm{GJ}}(V^{\mathrm{go}}_{N_{\mathrm{go}}-1}-V^{\mathrm{go}}_{N_{\mathrm{go}}})
 & \ (\mathrm{for} \quad i = N_\mathrm{go}),\end{array} \right.
 \label{J_i}
 \end{eqnarray}
where $V^{\mathrm{go}}_i$ is the membrane potential of the $i$th Golgi cell, $R^{\mathrm{go}}_i$ is the recovery variable representing the ion channel activity of the $i$th Golgi cell, $\mu$ is the parameter of the $\mu$-model, $I^{\mathrm{go, tonic}} $ is the common tonic input to all cells, $J^{\mathrm{go}}_i$ is the diffusion current into the $i$th Golgi cell from the neighboring cells conducted through the gap junctions, $g_{\mathrm{GJ}}$ is the conductance of a gap junction, $I^{\mathrm{go, input}}_i$ is the input signal to the reservoir described below, and $N_{\mathrm{go}}$ is the number of Golgi cells. The model only differs from that of the former studies (\cite{FUJII2004151,Tsuda2004, Tadokoro2011}) in that the input signal $I^{\mathrm{go, input}}_i$ is incorporated in Eq.~(\ref{V_i}). In the $\mu$-model, both the units of time and the variables are arbitrary. We use milliseconds as the unit of time for convenience. We use parameters $\mu = 1.7,\ g_{\mathrm{GJ}} =0.08$ (typical parameter set showing chaotic dynamics) except in Figs.~\ref{RMSEonGJ},~\ref{RMSEongMu} where the dependency of the dynamics on these parameters is studied, and in Figs.~\ref{MultipleFreq}, \ref{WithExcitatory} where $g_{\mathrm{GJ}}=0$ is used. 
For the parameter $I^{\mathrm{go, tonic}}$, we use $I^{\mathrm{go, tonic}} = 0.004$ other than in a simulation shown in Fig.~\ref{Sierpinski}(b) and (c).
 
We restrict the form of the input signal $\bm{I}^{\mathrm{go, input}} \in  \mathbb{R}^{N_{\mathrm{go}}}$ to an instantaneous pulse as follows:
\begin{eqnarray}
\bm{I}^{\mathrm{go, input}}&=& \bm{x}^{\mathrm{in}} \delta \left( t - t_1\right),
\label{xin}
\end{eqnarray}
 where $\delta \left( t \right)$ is the Dirac delta function, $\bm{x}^{\mathrm{in},i} \in  \mathbb{R}^{N_{\mathrm{go}}}$ is the ${N_{\mathrm{go}}}$-dimensional vector representing the amplitude of the input, $t_1$ is the time when the input is given to the reservoir. Practically, giving an input pulse is done by setting $\bm{V}^{\mathrm{go}} \rightarrow \bm{V}^{\mathrm{go}} + \bm{x}^{\mathrm{in}}$ at time $t_1$, where $\bm{V}^{\mathrm{go}} = (V^{\mathrm{go}}_i) \in \mathbb{R}^{N_{\mathrm{go}}}$ is the vector representation of the membrane potentials. In Sec. \ref{subsection:POT}, a series of two different pulses are given to the system. In this case, the input signal $\bm{I}^{\mathrm{go, input}}$ is described as follows: 
\begin{eqnarray}
\bm{I}^{\mathrm{go, input}}&=& \bm{x}^{\mathrm{in},1} \delta \left( t - t_1\right)+ \bm{x}^{\mathrm{in},2} \delta \left( t - t_2\right),
\end{eqnarray}
 where $t_1$ and $t_2$ are the times when the input pulses are given to the reservoir.

The model incorporating the granule cells shown in Fig.~\ref{arch}(b) is described as follows:
\begin{eqnarray}
\frac{\mathrm{d}V^{\mathrm{gr}}_i}{\mathrm{d}t} 
&=&
- R_i^{\mathrm{gr}} 
- \mu \cdot \left(V^{\mathrm{gr}}_i \right) ^2
(V^{\mathrm{gr}}_i \mathrm{-\frac{3}{2}})
+ I^{\mathrm{gr, tonic}} 
+I_i^{\mathrm{ei}} 
+I_{i}^{\mathrm{gr}, \mathrm{input}}
,
 \label{Vgra_i}
\\ 
\frac{\mathrm{d}R^{\mathrm{gr}}_{i}}{\mathrm{d}t} &=& - R_i^{\mathrm{gr}} - \mu \cdot \left( V^{\mathrm{gr}}_{i} \right)^2
,
 \label{Rgra_i}
\\ \noindent
\frac{\mathrm{d}V^{\mathrm{go}}_i}{\mathrm{d}t} 
&=&
- R_i^{\mathrm{go}} 
- \mu \left(V^{\mathrm{go}}_i \right) ^2(V^{\mathrm{go}}_i \mathrm{-\frac{3}{2}})
+ J^{\mathrm{go}}_i 
+I^{\mathrm{go, tonic}} 
+I_i^{\mathrm{ie}} 
+I_{i}^{\mathrm{go}, \mathrm{input}}
,
 \label{Vgo_i}
\\ 
\frac{\mathrm{d}R^{\mathrm{go}}_{i}}{\mathrm{d}t} &=& - R_i^{\mathrm{go}} - \mu \left( V^{\mathrm{go}}_{i} \right)^2,
 \label{Rgo_i}
 \end{eqnarray}
where $V^{\mathrm{gr}}_i$ is the membrane potential of the $i$th granule cell, $R^{\mathrm{gr}}_i$ is the recovery variable representing the ion channel activity of the $i$th granule cell,
$I^{\mathrm{gr, tonic}}$ is the common tonic input to the granule cells, and $I^{\mathrm{gr, input}}_i$ is the input signal to the reservoir projecting to the $i$th granule cell. The diffusion currents are described as follows:
 \begin{eqnarray}
J^{\mathrm{go}}_i&=& \left\{ \begin{array}{ll} 
g_{\mathrm{GJ}}(V^{\mathrm{go}}_{2}-V^{\mathrm{go}}_{1})
 & \ (\mathrm{for} \quad i = 1), \\
g_{\mathrm{GJ}}(V^{\mathrm{go}}_{i+1}+V^{\mathrm{go}}_{i-1}-2V^{\mathrm{go}}_{i})
 & \ (\mathrm{for} \quad i = 2,...,N_{\mathrm{go}}-1),\\
g_{\mathrm{GJ}}(V^{\mathrm{go}}_{N_{\mathrm{go}}-1}-V^{\mathrm{go}}_{N_{\mathrm{go}}})
 & \ (\mathrm{for} \quad i = N_\mathrm{go}).\end{array} \right.
 \label{Jgo_i}
\end{eqnarray}
The currents $I_i^{\mathrm{ei}}, I_i^{\mathrm{ie}} $ are the currents caused by chemical synapses, each representing the inhibition of the $i$th granule cell by the Golgi cells and the excitation of the $i$th Golgi cell by the granule cells, respectively, described with the following equations:
\begin{eqnarray}
I_i^{\mathrm{ei}} &=& \sum_{j=1}^{N_{\mathrm{go}}} w^{\mathrm{ei}}_{ij} f(V^{\mathrm{go}}_j - \theta),
\\
I_i^{\mathrm{ie}} &=& \sum_{j=1}^{N_{\mathrm{gr}}} w^{\mathrm{ie}}_{ij} f(V^{\mathrm{gr}}_j - \theta),
\end{eqnarray}
where $w^{\mathrm{ei}}_{ij}$ is the strength of the synaptic connection from the $j$th Golgi cell to the $i$th granule cell, $w^{\mathrm{ie}}_{ij}$ is the strength of the synaptic connection from the $j$th granule cell to the $i$th Golgi cell, $\theta$ is a parameter defining the threshold above which each neuron can be regarded as emitting a spike, and $f(V)$ is an activation function. In this study, we use $\theta=0.7,\ f(V) = \frac{1}{1+\mathrm{exp}(-50V)}$. The synaptic strengths are determined using the following procedure. For each granule cell $i$, $n_{\mathrm{ei}}$ presynaptic Golgi cells are randomly chosen. Then, the strength is set at $w^{\mathrm{ei}}_{ij} = c^{\mathrm{ei}} /n_{\mathrm{ei}}$, if the $j$th Golgi cell is in the chosen group. The strength is set at $w_{ij}^{\mathrm{ei}} = 0$, if the $j$th Golgi cell is not in the chosen group. Similarly, for each Golgi cell $i$, $n_{\mathrm{ie}}$ presynaptic granule cells are randomly chosen. Then, the strength is set at $w_{ij}^{\mathrm{ie}} = c^{\mathrm{ie}} /n_{\mathrm{ie}}$, if the $j$th granule cell is in the chosen group. The strength is set at $w_{ij}^{\mathrm{ie}} = 0$, if the $j$th granule cell is not in the chosen group. The parameter values used are $\mu = 1.7,\ g_{\mathrm{GJ}} \in \left\{ 0, \ 0.08\right\},\ N_{\mathrm{gr}}=10^4,\ N_{\mathrm{go}}=10^2, \ n_{\mathrm{ei}}=4,\ n_{\mathrm{ie}}=10^2,\ c^{\mathrm{ie}}=-c^{\mathrm{ei}} = 0.2$. These parameters are determined such that the ratio of the numbers of granule cells and Golgi cells, $N_{\mathrm{gr}}/N_{\mathrm{go}}$, and the number of synapses each neuron receive, $n_{\mathrm{ei}},\ n_{\mathrm{ie}}$, are compatible with those described in the former study (\cite{DeSchutter}).

Neurons in a specific subset of the reservoir neurons are connected to the outputs, which we refer to as the {\it projecting neurons} hereafter. Let $V_i^{\mathrm{out}}$ be the membrane potential of the $i$th projecting neuron. In the simple model shown in Fig.~\ref{arch}(a), all the Golgi cells are the projecting neurons. Thus, $V_i^{\mathrm{out}} = V^{\mathrm{go}}_i$. In the model shown in Fig.~\ref{arch}(b), all (and only) the granule cells are the projecting neurons. Thus, $V_i^{\mathrm{out}} = V^{\mathrm{gr}}_i$. The $i$th output of the reservoir, $y_i$ is defined as follows:
\begin{eqnarray}
y_i &=& \sum_{j=1}^{N_\mathrm{out}} w_{ij}^{\mathrm{out}} V_j^{\mathrm{out}},
 \end{eqnarray}
where $N_\mathrm{out}$ is the number of projecting neurons in the reservoir, and $w_{ij}^{\mathrm{out}}$ is the synaptic weight of the connection from the $j$th projecting neuron in the reservoir to the $i$th output $y_i$. The vector $\bm{y} = (y_i) \in \mathbb{R}^{N_y}$ defines the instantaneous output of the reservoir machine, where $N_y$ is the number of outputs (the number of Purkinje cells considered). The output synaptic weight $w_{ij}^{\mathrm{out}}$ is time independent, and its value is modified only in the batch learning procedure described below.

\subsection{Learning of the readout connection}

Let $\bm{y}^{\mathrm{target}} \in \mathbb{R}^{N_y}$ be the target pattern consisting of $N_y$-dimensional time series defined over a time interval $\left[t_{0}, \ t_{0}+T_{\mathrm{train}} \right]$, where $T_{\mathrm{train}}$ is the length of the time series. We determine the readout weight matrix $W^{\mathrm{out}} = \left[w_{ij}^{\mathrm{out}} \right] \in \mathbb{R}^{N_{y} \times N_{\mathrm{out}}}$ to minimize the following residual value:
\begin{eqnarray}
\int_{t_0}^{t_0 + T_{\mathrm{train}}} 
|\bm{y}^{\mathrm{target}}
-
\bm{y}|^2 
\mathrm{d}t
\noindent
= 
\int_{t_0}^{t_0 + T_{\mathrm{train}}} 
|
\bm{y}^{\mathrm{target}}
-
W^{\mathrm{out}} \bm{V}^{\mathrm{out}}
|^2 
\mathrm{d}t,
\label{optimization}
\end{eqnarray}
where $|\bm{x}|$ is the Euclidean norm of a vector $\bm{x}$. In practice, this is conducted by sampling both the output vectors and the target vectors with a small sampling interval $\Delta$. Let $\bm{\Omega} \ \in \mathbb{R}^{N_y \times K}$ be the discretized time series of the numerically integrated membrane potentials of the reservoir dynamics over the time interval $\left[t_0, t_0 + T_{\mathrm{train}} \right]$ as follows:
\begin{eqnarray}
\bm{\Omega} = \left( \bm{V}^{\mathrm{out}}({t_0}),\ \bm{V}^{\mathrm{out}}({t_0}+\Delta),\ \bm{V}^{\mathrm{out}}({t_0}+2\Delta),\ \cdots,\ \bm{V}^{\mathrm{out}}({t_0}+K\Delta)\right),\ 
\label{discretization}
\end{eqnarray}
where $K$ is the natural number satisfying $K\Delta \leq T_{\mathrm{train}} <(K+1)\Delta $, and $\bm{V}(t) = (V_i^{\mathrm{out}}(t)) \in \mathbb{R}^{N_{\mathrm{out}}}$ is the instantaneous membrane potentials of the projecting neurons. The target pattern matrix $\bm{Y}$ is also defined by the same discretization as follows:
\begin{eqnarray}
\bm{Y}^{\mathrm{target}} = \left(\bm{y}^{\mathrm{target}}({t_0}),\ \bm{y}^{\mathrm{target}}({t_0}+\Delta),\ \bm{y}^{\mathrm{target}}({t_0}+2\Delta),\ \cdots,\ \bm{y}^{\mathrm{target}}({t_0}+K\Delta)\right).
\end{eqnarray}
Then, the optimal readout weight matrix $\widehat{W}^{\mathrm{out}}$ is obtained by solving the following linear least square regression:
\begin{eqnarray}
\widehat{W}^{\mathrm{out}} = 
\argmin
\ 
|\bm{Y}^{\mathrm{target}}
-
W^{\mathrm{out}} \bm{\Omega} |_{\mathrm{fro}}^2
.
\label{LSR}
 \end{eqnarray}
where $|W|_{\mathrm{fro}}$ is the Frobenius norm of a matrix $W$. 

\subsection{Evaluation of model performance}
In order to evaluate the performance of the model, we use normalized root mean square error normalized by that of the target pattern (nRMSE), as follows:
\begin{eqnarray}
\mathrm{nRMSE} = 
\left(
\frac{
\displaystyle \int_{t_0}^{t_0 + T_{\mathrm{train}}} (
|
\bm{y}^{\mathrm{target}}
-
\widehat{W}^{\mathrm{out}} \bm{V}^{\mathrm{out}}
|^2 
)
\mathrm{d}t
}{
\displaystyle \int_{t_0}^{t_0 + T_{\mathrm{train}}} 
(
|
\bm{y}^{\mathrm{target}}
|^2 
)
\mathrm{d}t.
}
\right)^{\frac{1}{2}}
\end{eqnarray}
With the discretized time series, nRMSE is calculated as follows:
\begin{eqnarray}
\mathrm{nRMSE} = 
\dfrac{
 |\bm{Y}^{\mathrm{target}}
 -
 \widehat{W}^{\mathrm{out}} \bm{\Omega} |_{\mathrm{fro}}
 }{
 |\bm{Y}^{\mathrm{target}}} |_{\mathrm{fro}
 }.
\end{eqnarray}

\subsection{Lyapunov dimension}
For the simple model (Fig.~\ref{arch}(a)) under no dynamical input, we characterized the strength of chaotic activity of the reservoir with the Lyapunov dimension (\cite{DL}). First, we calculate the Lyapunov spectrum by the standard method with continuous Gram-Schmidt orthonormalization of the fundamental solutions to the linearized differential equation along the trajectory (\cite{ShimadaNagashima}). Then, let $\lambda_1 \geq \lambda_2 \geq \lambda_3 \cdots \geq \lambda_{2N_{\mathrm{go}}}$ be the Lyapunov exponents of the reservoir dynamics, and $k$ be the maximal value of $j$ such that $\sum_{i=1}^j \lambda_i \geq 0$, the Lyapunov dimension of the system is defined as follows:
\begin{eqnarray}
D_{\mathrm{L}} = k + \dfrac{\sum_{i=1}^k \lambda_i}{|\lambda_{k+1}|}.
\end{eqnarray}
In this study, the Lyapunov dimension of the system is calculated under no external input to the system, where the system can be regarded as an autonomous dynamical system. Because we restrict the form of the input to the system as an instantaneous pulse (Eq.~(\ref{xin}), the property of the reservoir without any external input characterizes the system's response to the external input.

\subsection{Similarity index}
It is commonly assumed that the cerebellar granular layer is able to exhibit activity specific to the passage of time in response to an input (\cite{Buonomano1994,YAMAZAKI2007290}). In order to evaluate the specificity of the instantaneous activity of the reservoir with respect to the passage of time and external input, we use the similarity index between different states of the reservoir. Let $\bm{V}^{\mathrm{out}, (1)}(t)$ and $\bm{V}^{\mathrm{out}, (2)}(t)$ be two different time series of the projecting neurons of the reservoir generated with two different external inputs. We use the following correlation function between the instantaneous values of $\bm{V}^{\mathrm{out}, (1)}(t)$ and $\bm{V}^{\mathrm{out}, (2)}(t)$ at two time points $t_1$ and $t_2$, which we call the similarity index:
\begin{eqnarray}
C^{(1), (2)}(t_1, t_2) = 
\dfrac{
  \sum_{i=1}^{N_\mathrm{out}} (V^{\mathrm{out}, (1)}_i(t_1) -\langle V^{\mathrm{out}, (1)}(t_1) \rangle)
  (V^{\mathrm{out}, (2)}_i(t_2) -\langle V^{\mathrm{out}, (2)}(t_2) \rangle)
  }{
  \sqrt{\sum_{i=1}^{N_\mathrm{out}} (V^{\mathrm{out}, (1)}_i(t_1) -\langle V^{\mathrm{out}, (1)}(t_1) \rangle)^2} \sqrt{\sum_{i=1}^{N_\mathrm{out}} (V^{\mathrm{out}, (2)}_i(t_2) -\langle V^{\mathrm{out}, (2)}(t_2) \rangle)^2}
  },
  \label{C}
\end{eqnarray}
where $\langle V^{\mathrm{out}, (1)}(t) \rangle$ is the mean membrane potential at time $t$ as follows:

\begin{eqnarray}
\langle V^{\mathrm{out}, (1)}(t) \rangle
=
\dfrac{1}{{N_{\mathrm{out}}}}\sum_{i=1}^{N_{\mathrm{out}}} V^{\mathrm{out}, (1)}_i(t).
\end{eqnarray}

\subsection{Numerical calculation}
The numerical simulations were conducted using the {\it ode45} function in Matlab R2019a (MathWorks Inc., Natick, MA, USA). The obtained time series of the reservoir states were further discretized with time step $\Delta = 0.1$ using the {\it interp1} function (Eq.~(\ref{discretization})). The learning procedure to obtain the optimal readout weight matrix $\widehat{W}^{\mathrm{out}}$ was conducted by solving a linear least squares regression in Eq.~(\ref{LSR}) using the Matlab function {\it mldivide}.

\section{Results}

\begin{figure}[h]
 \includegraphics[]{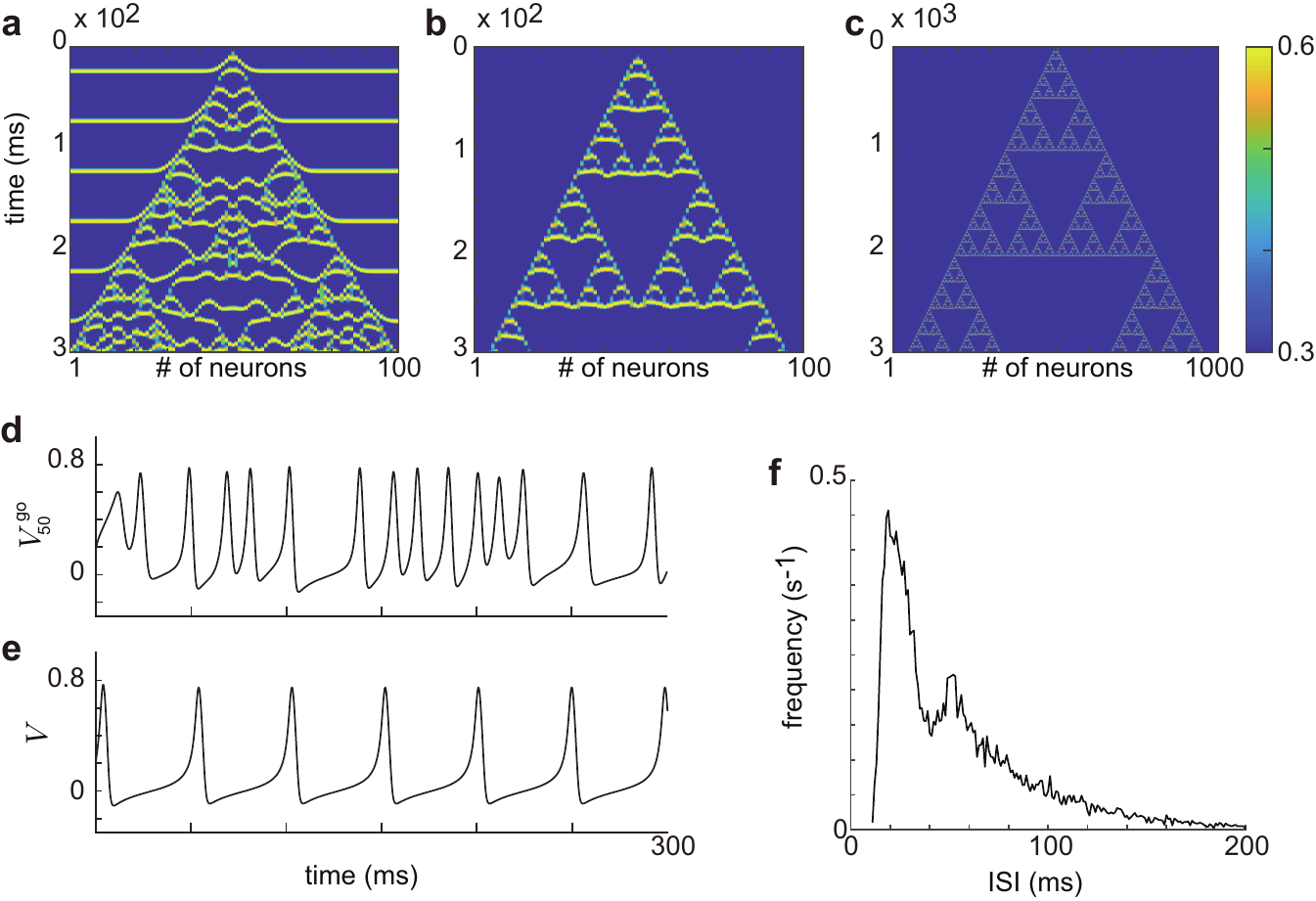}
  \caption{The chaotic dynamics induced by gap junctions in the simple reservoir consist of a one-dimensional chain (Eqs.~(\ref{V_i}-\ref{J_i}), Fig.~\ref{arch}(a)). (a) The dynamics of the model with a small positive tonic input, $I^{\mathrm{go, tonic}} = 0.004$. The colormap shows the dynamics of the membrane potentials, $\bm{V}^{\mathrm{go}}$. The parameter values are $N_{\mathrm{go}}=100,\ \mu = 1.7, \ g_{\mathrm{GJ}}= 0.08, \ I=0.004$. (b-c) A Sierpinski gasket appears under small tonic inhibition, $I^{\mathrm{go, tonic}} = -0.00095$. The number of neurons is $N=100$ in (b) and $N_{\mathrm{go}}=1000$ in (c). (d) The time evolution of the membrane potentials of the $50$th neuron in (a). (e) Periodic activity of an isolated $\mu$-model neuron under a small positive tonic input, $I^{\mathrm{go, tonic}} = 0.004$. (f) Broad distribution of the ISI of a neuron in chaotic dynamics. Data of the $50$th neuron in (a) for $1 \times 10^{3}$ seconds is shown. The bin width is $1$ ms.}
 \label{Sierpinski}
\end{figure}
\subsection{Broad distribution of the interspike interval caused by gap junctions}
First, we show that introducing diffusion coupling causes chaotic dynamics, which in turn results in a broad distribution of interspike interval (ISI) of a neuron in the model. Figure~\ref{Sierpinski} shows examples of the evolution of the simple reservoir (Fig.~\ref{arch}(a)) described by Eqs.~(\ref{V_i}-\ref{J_i}). Namely, the model network is a one-dimensional chain of neurons coupled to nearest neighbors with gap junctions and does not consist of any connections via chemical synapses. Figure~\ref{Sierpinski}(a) illustrates an evolution of the membrane potentials ($N_{\mathrm{go}}=100$) under a positive tonic input, $I^{\mathrm{go,tonic}}= 0.004$. An external perturbation at $t=0$ is given with $\bm{x}^{\mathrm{in}}$ in Eq.~(\ref{xin}) as $\bm{x}^{\mathrm{in}}=0.2\bm{\mathrm{e}}_{50}$, to the all-synchronized state, where $\bm{\mathrm{e}}_{50}$ is the 50th standard basis. Namely, the membrane potential of the neuron in the center (50th neuron) is set as $V^{\mathrm{go}}_{50} \rightarrow V^{\mathrm{go}}_{50} + 0.2$. The network shows chaotic activity induced by diffusion coupling, as reported previously (\cite{Tsuda2004, Tadokoro2011}). Periodic synchronous activity is observed before the propagation of chaotic dynamics is elicited by an external perturbation. The behavior of the cell in the center of the network is shown in Fig.~\ref{Sierpinski}(d). The activity is quite different from an isolated $\mu$-model neuron with the same parameter, as shown in Fig.~\ref{Sierpinski}(e) ($I^{\mathrm{go,tonic}}= 0.004, \ \mu=1.7$). An isolated $\mu$-model shows a saddle-node bifurcation at $I^{\mathrm{go, tonic}} =0$, and it has a periodic spiking activity with $I^{\mathrm{go, tonic}} >0$ and stable fixed point at resting potential with $I^{\mathrm{go, tonic}} <0$ (\cite{Tsuda2004}). The ISI of the neuron in the center of the network (cell $\# 50$) in Fig.~\ref{Sierpinski}(a) is calculated for the subsequent $1 \times 10^{3}$ seconds (Fig.~\ref{Sierpinski}(f)) by regarding the neuron emitting a spike when crossing the threshold $\theta=0.7$ from negative to positive. As visually evident in Figs.~\ref{Sierpinski}(a), (d), the ISI of this neuron shows a broad distribution over a wide range of periods, which is quite different from the periodic spiking activity without diffusion coupling shown in Fig.~\ref{Sierpinski}(e). The distribution has a small peak at around $50$ ms, which is close to the period of the isolated single neuron (without the effect of the gap junction) shown in Fig.~\ref{Sierpinski}(e). Interestingly, the spatiotemporal pattern of the membrane potentials shows the fractal known as the Sierpinski gasket (\cite{Mandelbrot}) under a small negative tonic input, $I^{\mathrm{go,tonic}}= -0.00095$ (Fig.~\ref{Sierpinski}(b)). The spatiotemporal pattern at larger scale ($N_{\mathrm{go}}=1000$) shown in Fig.~\ref{Sierpinski}(c) clearly depicts self-similarity of the spatiotemporal pattern. Namely, the spatiotemporal pattern has no characteristic scale. The spatiotemporal pattern with positive tonic input (Fig.~\ref{Sierpinski}(a)) could be interpreted as a ruined pattern of the Sierpinski gasket (Fig.~\ref{Sierpinski}(b)), thus inheriting the fractal’s property of scale invariance over multiple time scales (i.e., broad distribution of ISI).
\subsection{Gap junctions in the reservoir realize producing a target pattern with a broad range of frequencies.}
Next, we evaluate the effect of introducing gap junctions on the expressivity of the reservoir. More specifically, we examine how closely the model can output a sinusoidal temporal pattern with various temporal frequencies. Namely, we use the following sinusoidal wave (a scalar function of time $t$) as the target pattern $\bm{y}^\mathrm{target}$ described in Eq.~(\ref{optimization}):
\begin{eqnarray}
f(t; T^\mathrm{wave})= 
\sin{ \left( \dfrac{2 \pi t}{T^{\mathrm{wave}}}\right)},
\end{eqnarray}
where $T^{\mathrm{wave}}$ is the period of the sinusoidal wave. 
We quantify the dependency of the following nRMSE value on the period of the target sinusoidal wave, $T^{\mathrm{wave}}$, and on the model parameters:
\begin{eqnarray}
\mathrm{nRMSE}(T^{\mathrm{wave}})
=
\left(
\frac{
\mathrm{RSS}(T^{\mathrm{wave}})
}{
\int_{t_0}^{t_0 + T_{\mathrm{train}}} 
\left|f(t; T^\mathrm{wave})\right|^2
\mathrm{d}t
}
\right)^{\frac{1}{2}}
,
\label{nRMSEwave}
\end{eqnarray}
where $\mathrm{RSS}(T^{\mathrm{wave}})$ is the value of the objective function described in Eq.~(\ref{optimization}), which depends on $T^{\mathrm{wave}}$.
The spectrum of the value $\mathrm{nRMSE}(T^{\mathrm{wave}})$ over various $T^{\mathrm{wave}}$ gives a way to evaluate the model's ability to output a more general complex temporal pattern that contains various frequency components. Suppose there is a target pattern, 
$y^{\mathrm{target}}$ that is composed of a linear superposition of various sinusoidal waves as follows:
\begin{eqnarray}
y^{\mathrm{target}}=\sum_{k=1}^{N^{\mathrm{wave}}} 
c_k
f(t; T_k^\mathrm{wave}),
\label{linearSum}
\end{eqnarray}
where $N^{\mathrm{wave}}$ is the number of different sinusoidal waves that compose the target pattern, $T_k^\mathrm{wave}$ is the period of the $k$th sinusoidal pattern. Let $\mathrm{nRMSE^{\mathrm{target}}}$ be the optimal nRMSE values for $y^{\mathrm{target}}$ that minimizes the objective function described in Eq.~(\ref{optimization}). Then, from the triangle inequality, the following inequality holds:
\begin{eqnarray}
\left(\mathrm{nRMSE}^{\mathrm{target}}\right)^2 
\leq
\sum_{k=1}^{N^{\mathrm{wave}}} c_k^2
\left(
\mathrm{nRMSE}(T_k^\mathrm{wave})
\right)^2.
\label{ineq}
\end{eqnarray}
Thus, the right hand side of Eq. \ref{ineq} gives the upper limit of the nRMSE value for the target pattern.

\begin{figure}[H]
 \includegraphics[]{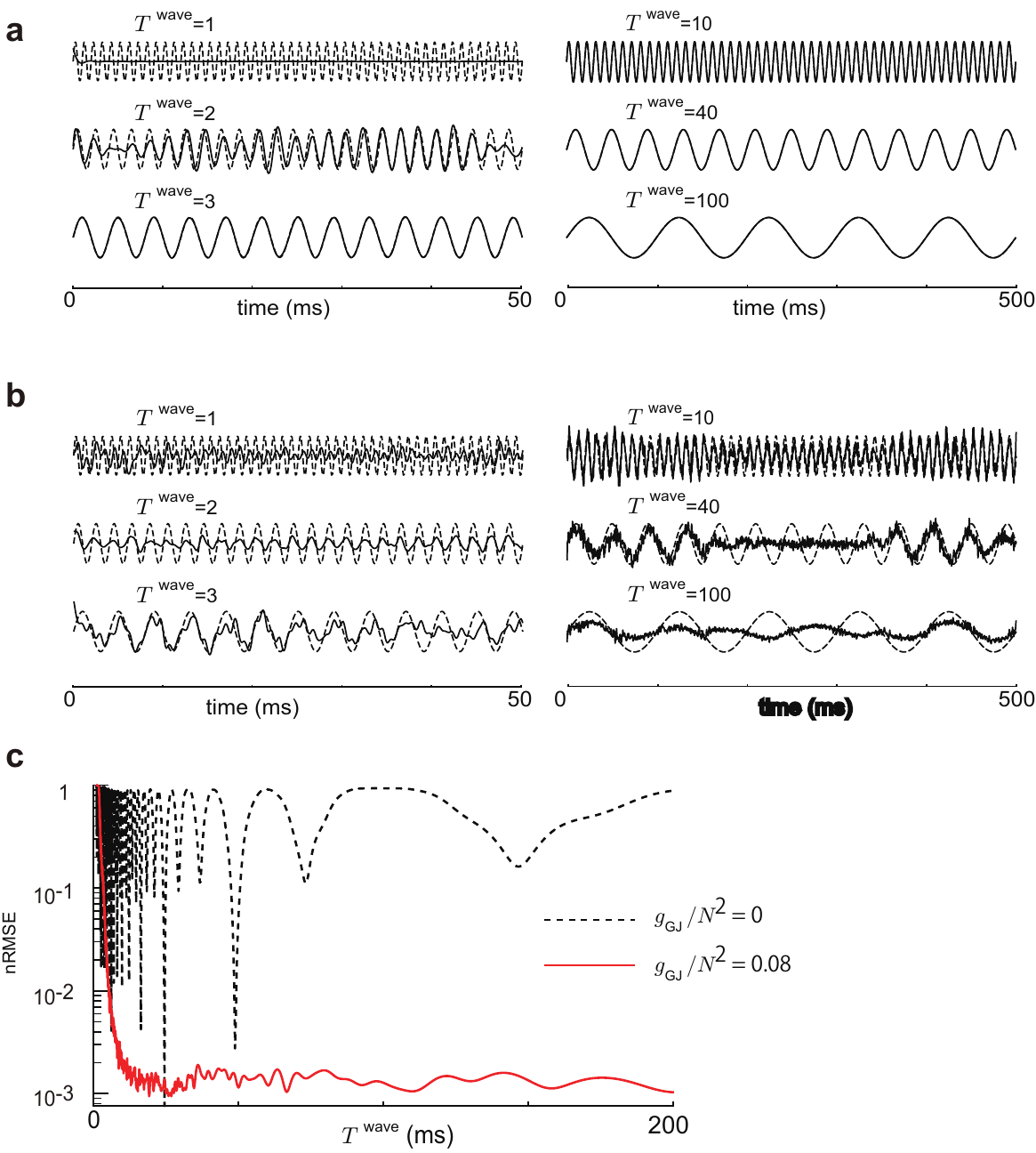}
  \caption{Evaluation of the expressivity in the frequency domain. (a) The readout weights are determined separately for each target sinusoidal pattern with a different period, $T^{\mathrm{wave}}$. Each dashed line indicates the target sinusoidal pattern and each solid line indicates the fitted output of the model. The parameters are $T_{\mathrm{train}} = 500 \ \mathrm{ms}, \ \mu = 1.7, \ g_{\mathrm{GJ}}=0.08, \ N_{\mathrm{go}} = 500, \ I^{\mathrm{go, tonic}} = 0.004, \ \Delta = 0.1$. Note that the spike width of a neuron is $\sim$ 5--8 ms, as shown in Fig.~\ref{Sierpinski}(b) and (c). Data only at $t = [0 \ 50]$ are shown for $T^{\mathrm{wave}} \leq 3$, though fitting is done for $t = [0 \ 500]$. (b) Without diffusion through the gap junctions ($g_{\mathrm{GJ}}=0$). Each solid line indicates the fitted output of the model. (c) The nRMSE dependency on the period of the target pattern, with gap junctions ($g_{\mathrm{GJ}}=0.08$) (solid line), and without gap junctions ($g_{\mathrm{GJ}}=0$) (dashed line).}
 \label{MultipleFreq}
\end{figure}
To evaluate nRMSE$(T^\mathrm{wave})$ (Eq.~(\ref{nRMSEwave})), the following analysis is conducted. Firstly, a time evolution over $\left[t_{0}, \ t_{0}+T_{\mathrm{train}} \right]$ of the reservoir with a random initial input at $t=0$ is numerically generated. Then, we use various sinusoidal waves (scalar function of time) over the same time interval $\left[t_{0}, \ t_{0}+T_{\mathrm{train}} \right]$ as the target patterns. The readout connection is fitted to match each sinusoidal wave separately by the linear least squares. The dashed lines in Fig.~\ref{MultipleFreq}(a) indicate the target sinusoidal patterns and the solid lines indicate the fitted outputs of the model. The difference between the target patterns and the model outputs are visually not detectable when $T^{\mathrm{wave}}\geq 3$ ms. Note that, as shown in Figs.~\ref{Sierpinski}(d) and (e), the width of a spike of the model neuron is $\sim$ 5--8 ms.  The results of the same analysis without gap junctions are shown in Fig.~\ref{MultipleFreq}(b). All settings of the analysis are the same as in Fig.~\ref{MultipleFreq}(a) except for two conditions. First, the conductance of the gap junction is changed from $g_{\mathrm{GJ}}=0.08$ to $g_{\mathrm{GJ}}=0$. Second, the initial state of the reservoir is generated by randomly shuffling the phases of neurons. This is because, with $g_{\mathrm{GJ}}=0$, all neurons behave as parallel isolated neurons with a shared identical period. Thus, the distribution of the phase of the neurons is time invariant, and biased distribution of the phase of the neurons should be disadvantageous in producing the sinusoidal target patterns. The precision of the model output drastically decreases compared to the condition with diffusion coupling through gap junctions (Fig.~\ref{MultipleFreq}(b)). The nRMSE values for both conditions are shown in Fig.~\ref{MultipleFreq}(c). The nRMSE takes very small values over a wide range of the period of the target pattern when the gap junctions are incorporated in the model, whereas the nRMSE takes small values only at some specific range of the period if the model lacks the gap junctions. This result suggests that incorporating diffusion coupling in a reservoir enhances the expressivity of the output that the network can generate.

\subsection{Inverse correlation of the Lyapunov dimension and {\rm nRMSE}}

\begin{figure}[H]
 \begin{minipage}[c]{0.45\textwidth}
 \begin{center}
  \includegraphics[]{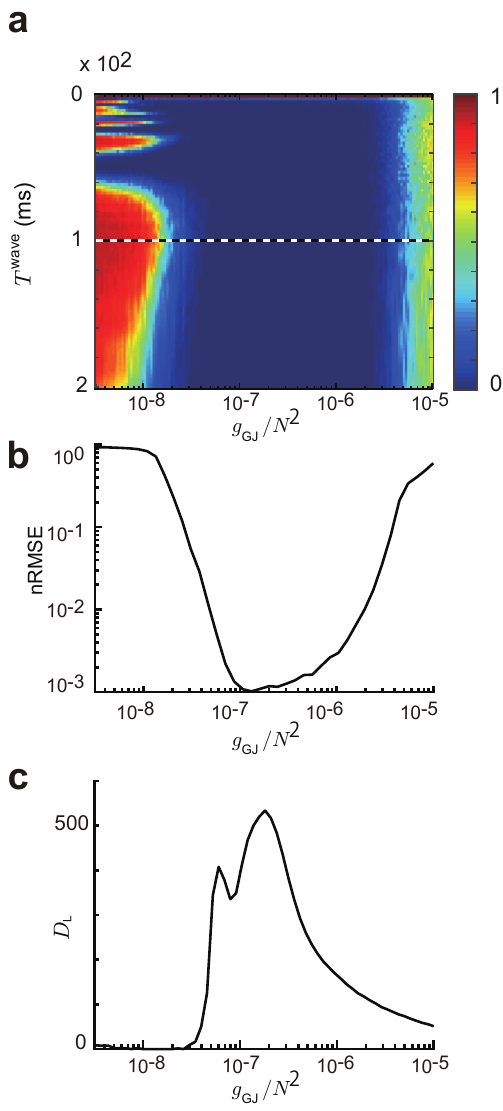}
  \caption{The dependency of the nRMSE on the period of the target pattern, $T^{\mathrm{wave}}$, and the strength of the gap junction, $g_{\mathrm{GJ}}/N_{\mathrm{go}}^2$.
  (a) The nRMSE is illustrated over various values of $T^{\mathrm{wave}}$ and $g_{\mathrm{GJ}}/N_{\mathrm{go}}^2$. The parameter values are $N_{\mathrm{go}} = 500, \ T_{\mathrm{train}} = 500 \ \mathrm{ms}, \ \mu = 1.7, \ I^{\mathrm{go, tonic}} = 0.004, \ \Delta = 0.1$.
  (b) The nRMSE value for a target pattern with $T^{\mathrm{wave}}=100 \ \mathrm{ms}$ plotted against $g_{\mathrm{GJ}}/ N_{\mathrm{go}}^2$. 
  (c) The Lyapunov dimension plotted against $g_{\mathrm{GJ}}/ N_{\mathrm{go}}^2$.} 
  \label{RMSEonGJ}
 \end{center}
  
 \end{minipage}
 \hfill
 \begin{minipage}[c]{0.45\textwidth}
 \begin{center}
 \includegraphics[]{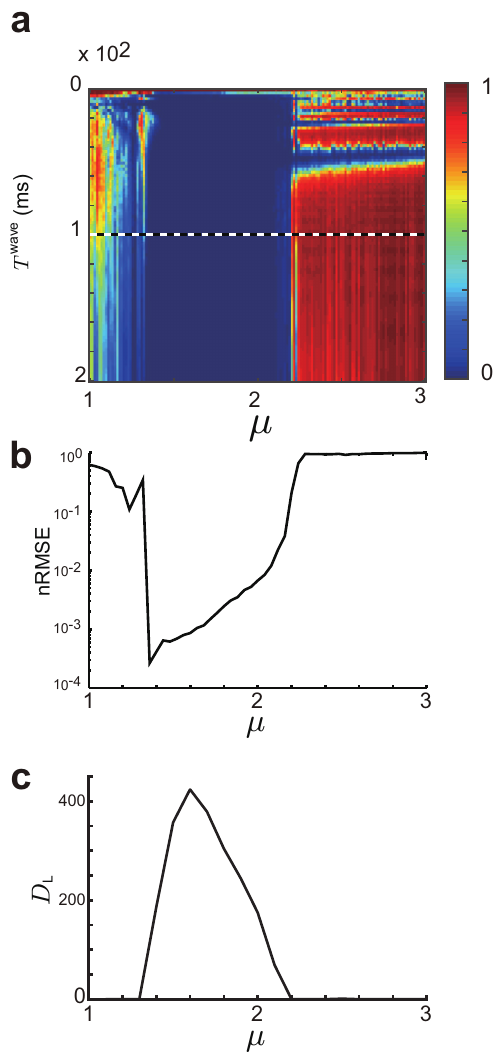}
  \caption{The dependency of the nRMSE on the period of the target pattern, $T^{\mathrm{wave}}$, and the parameter, $\mu$. 
  (a) The nRMSE is illustrated over various values of $T^{\mathrm{wave}}$ and $\mu$. The parameter values are $N_{\mathrm{go}} = 500, \ T_{\mathrm{train}} = 500 \ \mathrm{ms}, \ g_{\mathrm{GJ}} = 0.08, \ I^{\mathrm{go, tonic}} = 0.004, \ \Delta = 0.1$. 
  (b) The nRMSE value for a target pattern with $T^{\mathrm{wave}}=100 \ \mathrm{ms}$ plotted against $\mu$.
  (c) The Lyapunov dimension plotted against $\mu$.
   }
    \label{RMSEongMu}
  \end{center}
 
  \end{minipage}
\end{figure}

Figures \ref{RMSEonGJ} and \ref{RMSEongMu} illustrate the parameter dependency of nRMSE and the Lyapunov dimension. The color map in Fig.~\ref{RMSEonGJ}(a) depicts the dependency of the nRMSE on the strength of gap junction coupling, $g_{\mathrm{GJ}}/N_{\mathrm{go}}^2$, and the target wave period $T^{\mathrm{wave}}$. 
Figure~\ref{RMSEonGJ}(b) illustrates the nRMSE dependence on $g_{\mathrm{GJ}}/N_{\mathrm{go}}^2$
when the target pattern period is $T^{\mathrm{wave}} = 100$ ms.
We employ $g_{\mathrm{GJ}}/N_{\mathrm{go}}^2$ rather than $g_{\mathrm{GJ}}$ because different network models with different sizes, but a same value of $g_{\mathrm{GJ}}/N_{\mathrm{go}}^2$, behave qualitatively the same (\cite{Tadokoro2011}). This is because the model described with Eqs.~(\ref{V_i}-\ref{J_i}) can be regarded as a discretization of a partial differential equation of a continuous one-dimensional excitable media. Models with different network sizes, $N_{\mathrm{go}}$, and the same value of $g_{\mathrm{GJ}}/N_{\mathrm{go}}^2$ correspond to discretizations of the same partial differential equation with different spatial resolutions. Equivalently, different models with a same network size $N_{\mathrm{go}}^2$ and different $g_{\mathrm{GJ}}$ values show spatiotemporal patterns with different spatial scales proportional to $1/\sqrt{g_{\mathrm{GJ}}}$.
As illustrated in Figs.~\ref{RMSEonGJ}(a) and (b), the nRMSE takes small value at a wide but specific range of $g_{\mathrm{GJ}}$ ($10^{-7.5} \leq g_{\mathrm{GJ}} \leq 10^{-5.5}$). At the same time, the Lyapunov dimension, $D_{\mathrm{L}}$, takes a large value at the same range of $g_{\mathrm{GJ}}$ (Fig.~\ref{RMSEonGJ}(c)). The Lyapunov dimension characterizes the strength of chaos, and represents the degrees of freedom of the dynamics (\cite{DL}). Figure~\ref{RMSEongMu} illustrates the nRMSE dependency on the parameter $\mu$. Chaotic activity induced by diffusion coupling appears at a specific range of $\mu$, as shown in the positive Lyapunov dimension in Fig.~\ref{RMSEongMu}(c). As in the case of the dependency of the nRMSE value on $g_{\mathrm{GJ}}$ (Fig.~\ref{RMSEonGJ}), the nRMSE takes very small values when the dynamics is chaotic and the Lyapunov dimension is large. These results showing the inverse correlation between the Lyapunov dimension and nRMSE suggest that diffusion-induced chaotic dynamics enhances the complexity of the representation in the reservoir.

\subsection{The reservoir state represents the passage of time from a specific input}
\label{subsection:POT}

\begin{figure}[H]
  \begin{minipage}[c]{0.65\textwidth}
  \begin{center}
  \includegraphics[]{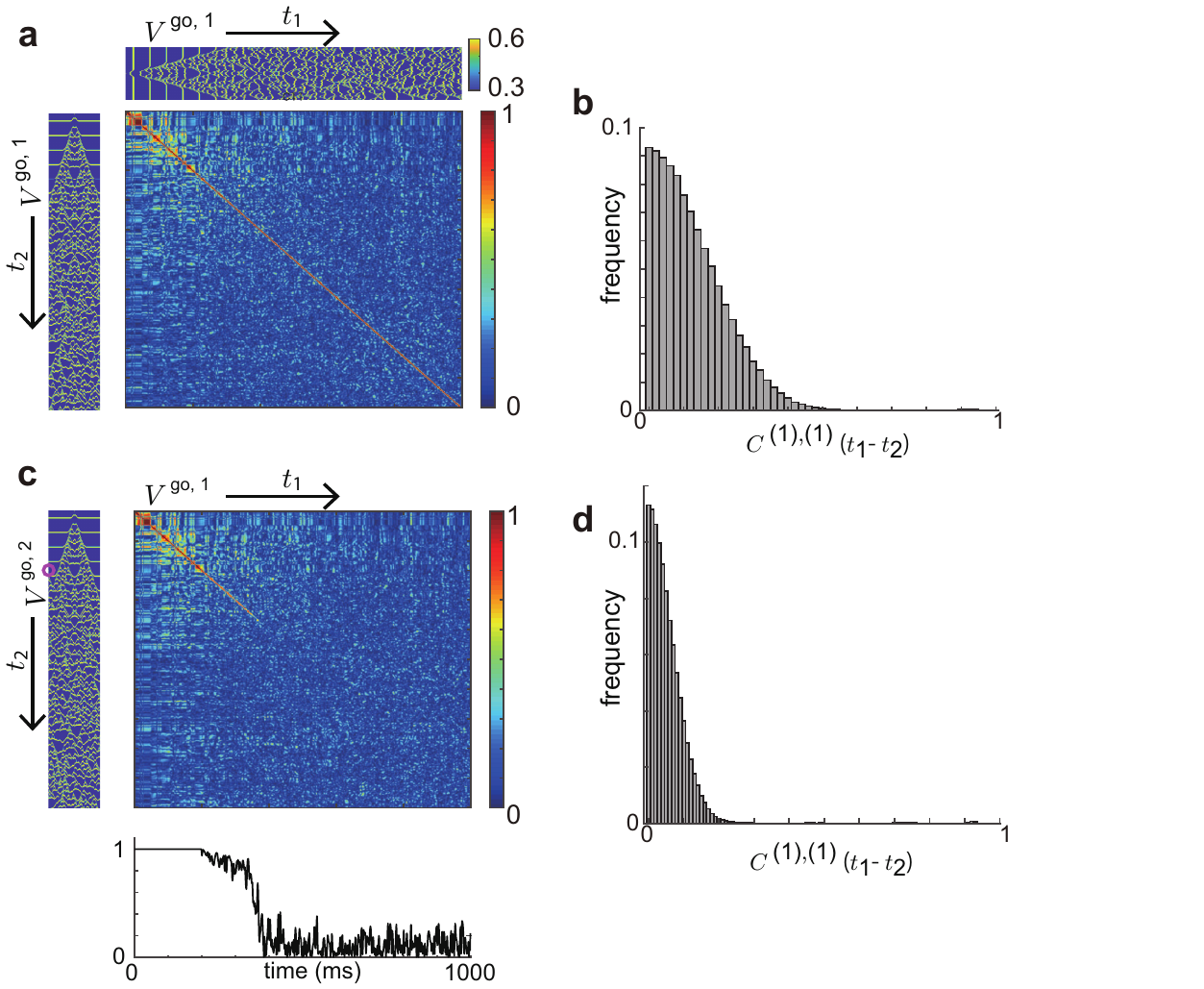}
    \end{center}
  \end{minipage}
  \begin{minipage}[c]{0.35\textwidth}
    \caption{The reservoir state is specific to the passage of time from a specific input. (a) Similarity indices $C^{(1), (1)}(t_1, t_2)$ calculated within a time series. The upper and left panels show the time series, $\bm{V}^{\mathrm{go}, 1}(t)$: the membrane potentials of reservoir neurons over 1000 ms. The neuron in the center (cell $\# 50$) is stimulated at $t=0$ as $V^{\mathrm{go}}_{50} \rightarrow V^{\mathrm{go}}_{50} + 0.5$. The parameter values are $\mu = 1.7, \ g_{\mathrm{GJ}}=0.08, \ N_{\mathrm{go}} = 100, \ I^{\mathrm{go, tonic}} = 0.004$. (b) Histogram of the upper triangular part of the matrix of similarity indices shown in (a). The bin width is $0.01$. Normalized with total counts. (c) The similarity indices calculated between two time series, $\bm{V}^{\mathrm{go}, 1}(t)$ and $\bm{V}^{\mathrm{go}, 2}(t)$. The time series in the left, $\bm{V}^{\mathrm{go}, 2}(t)$, evolves with the same initial condition as $\bm{V}^{\mathrm{go}, 1}(t)$, but an additional input pulse is given at the $100$th cell at $t = 200$ ms (shown with an open magenta circle). The lower panel shows the diagonal elements versus time. (d) Histogram of the upper triangular part of a matrix of similarity indices calculated with a model with $N_{\mathrm{go}} = 500$ cells. Normalized with total counts.}
 \label{Corr2time}
 \end{minipage}
\end{figure}

We evaluate the reservoir's ability to represent the passage of time with the similarity index (Eq.~(\ref{C})). The color map in Fig.~\ref{Corr2time}(a) shows similarity indices $C^{(1), (1)}(t_1, t_2)$ defined by Eq.~(\ref{C}), calculated within a time series, $\bm{V}^{\mathrm{go}, 1}(t)$, shown in the upper and left panels in Fig.~\ref{Corr2time}(a). An external perturbation is given at $t=0$ to an all-synchronized state as $V^{\mathrm{go}}_{50} \rightarrow V^{\mathrm{go}}_{50} + 0.5$. The length of the time series $\bm{V}^{\mathrm{go}, 1}(t)$ is 1000 ms, and it is discretized with a time step of $1$ ms to calculate the similarity indices, yielding a $1000 \times 1000$ matrix. Each element of the matrix corresponds to the correlation coefficient between the membrane potentials of the reservoir neurons at two different time points. Because $C^{(1), (1)}(t_1, t_2)$ is calculated within one time series, the diagonal elements are all 1. Figure~\ref{Corr2time}(b) shows the histogram of the values of upper triangular elements of the $C^{(1), (1)}(t_1, t_2)$ shown in panel (a). Most of the elements have smaller values than $\sim 0.4$. This suggests that the state of the system does not come back close to the same point in the phase space --close enough that the similarity index takes high value close to 1 -- within 1000 ms. In other words, the state of the reservoir activity has specificity to the passage of time. The distribution of the similarity index between different time points shifts towards even smaller values with larger network size, as shown in the case of $N_{\mathrm{go}} =500$ (Fig.~\ref{Corr2time}(d)). 
\\

The discriminative ability of the reservoir to different inputs is also important. Figure~\ref{Corr2time}(c) shows the similarity indices calculated between two time series, $\bm{V}^{\mathrm{go}, (1)}(t)$ and $\bm{V}^{\mathrm{go}, (2)}(t)$, that have slightly different inputs. The time series shown in the left panel, $\bm{V}^{\mathrm{go}, (2)}(t)$, evolves with the same initial condition as $\bm{V}^{\mathrm{go}, (1)}(t)$ shown in Fig.~\ref{Corr2time}(a), but an additional input pulse is given to the cell at the edge of the network at $t = 200$ as $V^{\mathrm{go}}_{1} \rightarrow V^{\mathrm{go}}_{1} + 0.5$ (shown with an open magenta circle). Because of the chaotic property of the dynamics, the orbit diverges from the original unperturbed orbit of $\bm{V}^{\mathrm{go}, (1)}(t)$, and the correlation between the two time series vanishes at around $t=350$ ms (Fig.~\ref{Corr2time}(c), lower panel). This can be explained by the sensitivity to the initial condition of chaotic dynamics. Thus, the reservoir's response to input has high specificity to the input.
\subsection{Generation of different activities for different inputs}
\begin{figure}[H]
\begin{center}
  \includegraphics[]{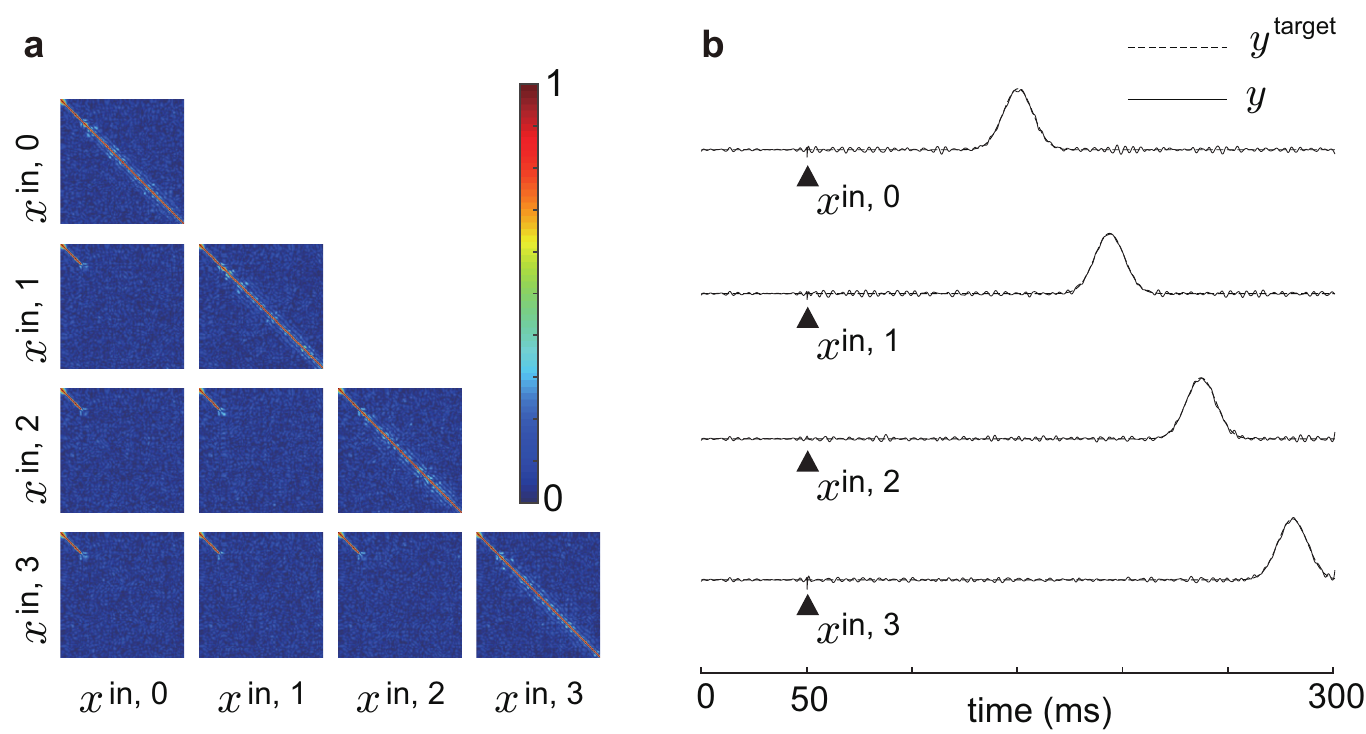}
\end{center}
    \caption{Different output patterns for different inputs acquired by the model with a same readout connection matrix. (a) Each heat maps show the similarity indices calculated between two different time series corresponding to two different inputs. The parameter values are $\mu = 1.7, \ g_{\mathrm{GJ}}=0.08, \ N_{\mathrm{go}} = 500, \ I^{\mathrm{go, tonic}}  = 0.004$. 
   (b) The time series with the initial inputs $\bm{x}^{\mathrm{in}, 0}, \ \bm{x}^{\mathrm{in}, 1}, \ \bm{x}^{\mathrm{in}, 2}, \ \bm{x}^{\mathrm{in}, 3}$ are fitted to different target patterns (dashed lines) with a shared readout connection. The output patterns are plotted with solid lines.
    }
 \label{Eyeblink}
\end{figure}

Next, we examine whether a model with a fixed readout weight matrix is actually able to generate different temporal patterns as outputs for different inputs. Firstly, we show a simulation of eyeblink conditioning: an extensively studied model of cerebellar dependent learning, which has been reproduced in several computational studies as well (\cite{Buonomano1994, BULLOCK19941101, Medina5516, YAMAZAKI2007290, LI201395, YAMAZAKI2013103}). We simulated a situation where the model outputs a specific time series with respect to a specific external input. In eyeblink conditioning, animals are able to acquire motor response with different timings to different types of tone stimuli (\cite{Kehoe1993, Green01052005}). For example, an animal can learn to elicit a motor reflex to a tone stimulus 200 ms after the stimulus onset if the pitch of the tone is 600-Hz, and 600 ms after the onset if its pitch is 1-kHz (\cite{Kehoe1993}). To reproduce this phenomenon, we consider the case where the model output, $y$, is a scalar function of time representing the eyeblink response, and train the model with multiple target patterns with its peaks at different latencies from the external input. Figure~\ref{Eyeblink}(a) shows similarity indices calculated between 
four time series of the reservoir state with a same initial condition and four different inputs $\bm{x}^{\mathrm{in}, 0}, \ \bm{x}^{\mathrm{in}, 1}, \ \bm{x}^{\mathrm{in}, 2}, \ \bm{x}^{\mathrm{in}, 3}$ given at time $t=50$. Four inputs, $\bm{x}^{\mathrm{in}, 0}, \ \bm{x}^{\mathrm{in}, 1}, \ \bm{x}^{\mathrm{in}, 2}, \ \bm{x}^{\mathrm{in}, 3}$ are generated from a multivariate Gaussian distribution $\mathcal{N}(\bm{0},\,(0.2)^{2}\bf{E})$, where $\bf{E}$ is the identity matrix. As shown in Fig.~\ref{Eyeblink}(a), the four spatiotemporal patterns with the inputs $\bm{x}^{\mathrm{in}, 0}, \ \bm{x}^{\mathrm{in}, 1}, \ \bm{x}^{\mathrm{in}, 2}, \ \bm{x}^{\mathrm{in, 3}}$ rapidly lose similarity among them. When the time series with the inputs $\bm{x}^{\mathrm{in}, 0}, \ \bm{x}^{\mathrm{in}, 1}, \ \bm{x}^{\mathrm{in}, 2}, \ \bm{x}^{\mathrm{in}, 3}$ are fitted to different target patterns (dashed lines in Fig.~\ref{Eyeblink}(b)) simultaneously, the model acquires different outputs assigned for each time series of the reservoir (solid lines).

\begin{figure}[H]
  \includegraphics[]{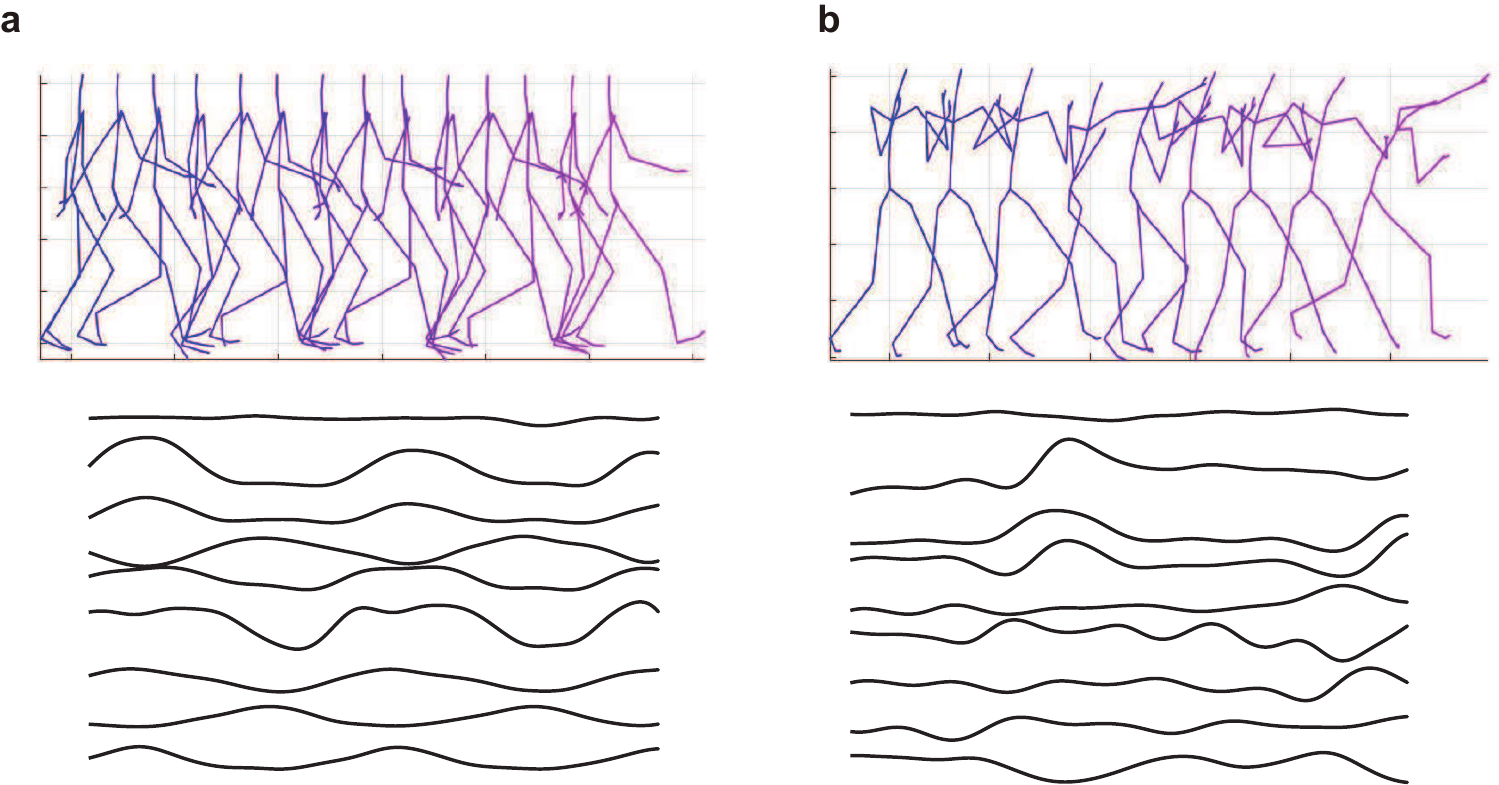}
    \caption{
    The network showing walking (a) and boxing (b) patterns for two different initial inputs. The upper panels show the human motion generated by the model output visualized using a skeleton, and the lower panels show the corresponding raw output signals of the model representing the angles of the bones. Only the representative nine out of 59 traces are shown for simplicity. The same readout connection matrix is used for both movements, walking and boxing. The motion capture data from a human are learned. The parameters are $\mu = 1.7, \ g_{\mathrm{GJ}}=0.08, \ N_{\mathrm{go}} = 500, \ I^{\mathrm{go, tonic}}  = 0.004$. Visualization was done using a toolbox, MATLAB Motion Capture Toolbox, at https://github.com/lawrennd/mocap. See also the supplementary video (walk.avi, box.avi).
    }
 \label{Walkbox}
\end{figure}

We also demonstrate the model's ability to generate two different human motions responding to two different inputs. Figure~\ref{Walkbox} shows the two outputs of a learned model with a fixed output matrix: walking (Fig.~\ref{Walkbox}(a)) and boxing (Fig.~\ref{Walkbox}(b)). The model is trained on two time series simultaneously, each for a specific input at time $t=0$. The motion capture data provided by the Carnegie Mellon University Motion Capture Library (MOCAP) (http://mocap.cs.cmu.edu/) were used. Datasets 08\_01.amc (walking) and 13\_17.amc (boxing) were used as the target output patterns. Three variables representing the spatial offset of the person are discarded from the 62 dimensional signal. Thus, the target patterns are the remaining 59-dimensional temporal patterns. The result demonstrates that the model is able to generate completely different temporal patterns with a fixed readout connection, when the initial condition is different. See also the supplementary video (walk.avi, box.avi).

\subsection{Incorporation of excitatory granular neurons}
\begin{figure}[H]
  \begin{minipage}[c]{0.65\textwidth}
  \begin{center}
  \includegraphics[]{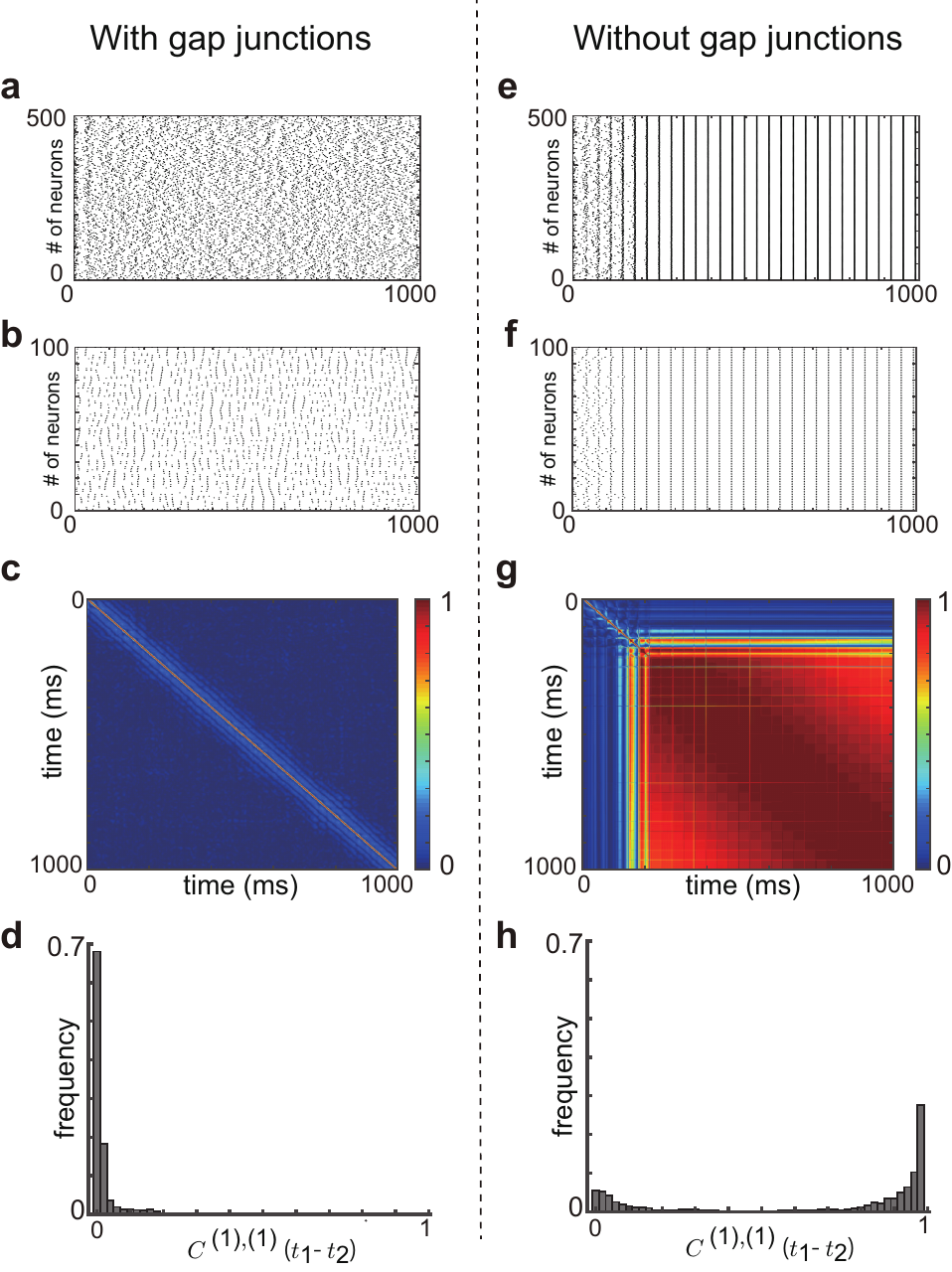}
    \end{center}
  \end{minipage}
  \begin{minipage}[c]{0.35\textwidth}
    \caption{The dynamics of the model incorporating the granule cells in the reservoir. (a-d) The model behavior with gap junctions under random initial input. The parameters are $\mu = 1.7, \ g_{\mathrm{GJ}}=0.08, \ N_{\mathrm{gr}} = 10^4, \ N_{\mathrm{go}} = 100, \ I^{\mathrm{gr, tonic}}  = 0.01, \ I^{\mathrm{go, tonic}}  = 0.004, \ \theta = 0.7$. (a) The spike raster plot of the granule cells. Only the 1-500th cells out of $N_{\mathrm{gr}} = 10^4$ cells are shown. (b) The spike raster plot of all the Golgi cells. (c) The similarity index calculated between instantaneous membrane potentials of the granule cells at different time points in the time series shown in (a). (d) Histogram of the upper triangular part of the matrix of similarity indices shown in (c). Normalized with total counts. (e-h) The model dynamics without gap junctions. All of the conditions of simulation and properties of the model are the same as those in (a), other than that parameter value $g_{\mathrm{GJ}}=0$ is used. (e) Spike raster plot showing the activities of the granule cells of a model without gap junctions. (f) The spike raster plot of all the Golgi cells. (g) The similarity indices showing high values over different time points because of the rapid relaxation to a synchronized activity. (h) Histogram of the upper triangular part of the matrix of similarity indices shown in (g). }
 \label{WithExcitatory}
\end{minipage}
\end{figure}

Lastly, we briefly confirm that the observed property of the simple model (Fig.~\ref{arch}(a)) can also be reproduced in the model incorporating the excitatory granule cells and the chemical synapses, shown in Fig.~\ref{arch}(b) (Eqs.~(\ref{Vgra_i})--(\ref{Jgo_i})). A model incorporating $10^4$ granule cells, $100$ Golgi cells, and the reciprocal connections between the granule cells and the Golgi cells with the chemical synapses is composed. We evaluate the similarity index using the membrane potentials of the granule cells because the projecting neurons of the real cerebellar granular layer to the Purkinje cells are the granule cells. Figure~\ref{WithExcitatory} shows examples of the dynamics of the model after a random initial input, 
with the gap junctions ($g_{\mathrm{GJ}}=0.08$, Figs.~\ref{WithExcitatory}(a)-(d)) and without gap junctions ($g_{\mathrm{GJ}}=0$, Figs.~\ref{WithExcitatory}(e)-(f)). Figures \ref{WithExcitatory}(a) and (b) show the raster plots of the spikes of the granule cells and the Golgi cells, respectively. Both the granule cells and the Golgi cells show irregular activity with no apparent repetitive pattern. The similarity index between two different time points within this time series verifies that the state of the granule cells, $\bm{V}^{\mathrm{gr}}$, is specific to time (Fig.~\ref{WithExcitatory} ~(c)). The histogram of the upper triangular elements of this matrix reveals the small similarity indices between two different time points (Fig.~\ref{WithExcitatory} ~(d)). The similarity index is distributed at far smaller values than that of the Golgi membrane potentials in the simple model (Figs.~\ref{Corr2time}(c) and (d)), presumably because of the large number of granule cells. Figures \ref{WithExcitatory} ~(e)--(h) show the result of same analysis with the parameter $g_{\mathrm{GJ}}=0$. All the other settings are identical. It is observed that, without diffusion coupling, the model dynamics converges to an all-synchronized periodic orbit by the interaction through the chemical synapses. The similarity indices (Fig.~\ref{WithExcitatory} ~(g)) and the distribution of non-diagonal elements clearly show that changing the parameter $g_{\mathrm{GJ}}$ to $0$ abolished the model's ability to represent the passage of time. These results suggests that the stable all-synchronized orbit exists in the system without gap junction, but it is destroyed by the chaotic dynamics if diffusion current through gap junctions exists.

\section{Discussion}

In the current study, we investigated the computational role of the gap junction between Golgi cells in the cerebellar granular layer. Specifically, we evaluated the computational performance of the model of the cerebellar cortex using a reservoir computing framework. First, we showed that introducing gap junctions in the model induces chaotic dynamics that enables the reservoir to output complex patterns containing a wide range of frequency components (Figs.~\ref{Sierpinski}-\ref{RMSEongMu}). Second, we showed that the chaotic dynamics has a long non-recursive time series that is capable of representing the passage of time (Fig.~\ref{Corr2time}). These properties of the chaotic dynamics realize the reservoir's ability to output the desired temporal patterns (Figs.~\ref{Eyeblink}, \ref{Walkbox}). Yamazaki and their colleagues have proposed a model with these abilities based on a different mechanism, i.e., the random connection by chemical synapses between granule cells and Golgi cells (\cite{YAMAZAKI2007290}). In the current study, we pointed out another possible mechanism (diffusion through gap junctions) that would be capable of reproducing the aforementioned abilities of the model. 
Because the gap junctions connect neighboring neurons, the connections realized by the gap junctions must inevitably be local rather than distant. Thus, the average number of neighboring cells a neuron can contact cannot be as large as the case of chemical synapses. The average degree of the network realized by gap junctions may therefore be small. In the literature, it has been argued that the desirable feature of a reservoir is that the connection should be sparse (\cite{jaeger:techreport2001}). This is in line with our hypothesis that diffusion coupling by the gap junction constitutes the reservoir in the cerebellum. On the other hand, one report showed that the small-worldness of the reservoir contributes to better performance (\cite{KAWAI201915}). In the granular layer of the real brain, the chemical synapses and the gap junctions may work in concert to realize preferable properties as a reservoir.

The ISI of a neuron in the chaotic dynamics induced by diffusion coupling through gap junctions shows a broad distribution over a wide range of periods, unlike that of an isolated neuron model (Fig.~\ref{Sierpinski}). Our result may explain the fact that a Golgi cell exhibits periodic activity in some experimental settings such as {\it in vitro} recordings (\cite{Forti, Solinas, Vervaeke2010}), but also shows an irregular activity with broad ISI distribution {\it in vivo} (\cite{Holtzman}). 

Reservoir computing has drawn considerable attention in recent years, as it is expected to be a suitable and powerful framework for processing temporal sequences. However, the desirable dynamical properties that the reservoir must have, and its proper implementation, have not been well characterized, and still remain an open question (\cite{Boyd,jaeger:techreport2001,Maass2002,EdgeOfChaos2,EdgeOfChaos,YILDIZ20121}). A variety of systems, including both physical systems and mathematical models, have been used as implementations of the reservoir (\cite{TANAKA2019100}). The current study investigated a reservoir consisting of a reaction-diffusion system (i.e., neurons coupled with gap junctions) and obtained results suggesting chaos in a reaction-diffusion system may contribute to the performance of the model. Further investigation of a reservoir machine using chaotic dynamics in a reaction-diffusion system may be an interesting direction for future study.

We found that the spatiotemporal pattern of the chaotic dynamics of $\mu$-models coupled with diffusion in a one-dimensional chain shows the Sierpinski gasket (Fig.~\ref{Sierpinski}). It is reported that some other nonlinear reaction-diffusion systems also shows the Sierpinski gasket (\cite{Hayase1998, Hayase2000}). The simple model in our current study (Fig.~\ref{arch}(a)) also belongs to the class of reaction-diffusion system, as it consists of a one-dimensional chain of the neurons with nearest neighbor connections. There maybe a common mathematical structure behind our model and these former studies. It is well known that the Sierpinski gasket appears in the spatiotemporal patterns of a cellular automaton (CA) such as Wolfram's Rule 90 (\cite{Wolfram}). Some previous studies attempted to implement a reservoir with a CA (\cite{Yilmaz14a, alej2018reservoir}). Mor\'{a}n et al. used CA as the reservoir and constructed a classifier for handwritten characters of the MNIST dataset (\cite{lecun-mnisthandwrittendigit-2010}), and compared the performance across the rules in CA. They reported that Wolfram's Rule 90 gives the best performance in the test data of the cross-validation (\cite{alej2018reservoir}). In the current study, we did not construct a classifier with our model. Thus, it is difficult to directly compare the current results with Mor\'{a}n and their colleagues' work. However, it should be an interesting direction to evaluate the performance of a classifier model with a reaction-diffusion system as the reservoir. Additionally, studies with CA may contribute to the elucidation of the cerebellar computation.

An important issue we did not consider in depth in the current study is the generalization ability of the model. Namely, the ability of the model to generate the same output from similar input. We showed that the chaotic dynamics realizes the specificity of the response to the input (Fig.~\ref{Corr2time}). This can be interpreted by the chaotic dynamics' high sensitivity to the initial condition. However, the high sensitivity to the initial condition may cause poor generalization ability, because a small noise or a deviance in the input signal grows rapidly over time. Thus, systematic evaluation of the generalization ability of the current model should be an important issue to be elucidated in the future study. The aforementioned study by Mor\'{a}n et al. showed that Wolfram's Rule 90, which shows the same Sierpinski gasket pattern as the current model we use, shows the best performance in the test data of cross-validation (i.e., it shows the highest generalization ability). In their study, the MNIST data is used as the initial state of the reservoir, and the spatiotemporal pattern of the evolution of CA over specific steps is used as the feature vector used for the subsequent classification task. Similarly, one could modify our current model so that the activity caused by the input decays within a specific time scale, before the small deviance in the input grows to the system size. This situation is actually similar in the real cerebellum because the cerebellum is believed to be able to maintain the input information for a fixed time, approximately $\sim 500$ ms (\cite{Thompson, Kotani}). It should be an interesting issue to elucidate the relationship to previously proposed properties of the reservoir such as the echo state property (\cite{jaeger:techreport2001, YILDIZ20121}) or the edge of chaos (\cite{EdgeOfChaos2, EdgeOfChaos}). Another important issue to be elucidated is the relationship between the strength of gap junctions and the generalization ability of the model. As shown in Fig.~\ref{RMSEonGJ}, the Lyapunov dimension of the system takes a large value at a specific range of the strength of gap junctions. Some previous studies pointed out the possibility that the strength of gap junctions changes the degrees of freedom of the network dynamics, which is crucial for the generalization ability (\cite{KAWATO2011791, Schweighofer2013, TOKUDA201758, Hoang542183}). Additionally, it should also be noted that cerebellar dependent motor learning requires repetitive training (\cite{Thompson}), which would help generalization. This is very different from hippocampal dependent learning, where an episode is learned one-shot.

In the last part of the Results section, we confirmed that the non-recursiveness of the system is inherited in the model incorporating the excitatory granule cells (Fig.~\ref{WithExcitatory}). Actually, the similarity index between different time points within a time series (Fig.~\ref{WithExcitatory}(d)) shows higher specificity to the passage of time than in the simple model consisting of only the Golgi cells (Fig.~\ref{Corr2time}(c)). This suggests that the large number of granular cells contributes to the specificity. In this study, we incorporated granular neurons with a population size only $10^2$ times larger than the Golgi cells ($N_{\mathrm{gr}} = 10^4, \ N_{\mathrm{go}} = 10^2$) because of the computational cost. However, the ratio of the number of the granule cells to the Golgi cells in the real brain is reported to be even larger, as much as 430 times (\cite{KORBO1993262}). The numerous granule cells may serve a role in multiplying the representational ability of the Golgi neurons.

In conclusion, we proposed the hypothesis that the massive gap junctions between the Golgi cells in the cerebellar granular layer contributes to expressivity by inducing chaotic dynamics. 

\section{Acknowledgements}
This work was supported by JSPS KAKENHI (Nos. 17K16365, 19K12235, 20H04258 and 20K19882). This study was also partially supported by the JST Strategic Basic Research Programs (Symbiotic Interaction: Creation and Development of Core Technologies Interfacing Human and Information Environments, CREST Grant Number JPMJCR17A4). This paper is based on results obtained from a project commissioned by the New Energy and Industrial Technology Development Organization (NEDO).

\section{Declaration of Conflict of Interest}
The authors declare that there is no conflict of interest.

\bibliography{Bibfile_ref}

\end{document}